\documentclass[aip,reprint,nofootinbib]{revtex4-1}
\usepackage{graphicx}
\usepackage{bm}

\begin{document}
\title{A New Look at Linear (Non-?) Symplectic Ion Beam Optics in Magnets}
\date{\today}
\author{C. Baumgarten}
\affiliation{Paul Scherrer Institute, Switzerland}
\email{christian.baumgarten@psi.ch}

\def\begeq{\begin{equation}}
\def\endeq{\end{equation}}
\def\begary{\begeq\begin{array}}
\def\endary{\end{array}\endeq}
\def\bmtx{\left(\begin{array}}
\def\emtx{\end{array}\right)}
\def\eps{\varepsilon}
\def\g{\gamma}
\def\y{\gamma}
\def\w{\omega}
\def\W{\Omega}
\def\s{\sigma}

\def\Exp#1{\exp\left(#1\right)}
\def\Log#1{\ln\left(#1\right)}
\def\Sinh#1{\sinh\left(#1\right)}
\def\Sin#1{\sin\left(#1\right)}
\def\Tanh#1{\tanh\left(#1\right)}
\def\Tan#1{\tan\left(#1\right)}
\def\Cos#1{\cos\left(#1\right)}
\def\Cosh#1{\cosh\left(#1\right)}

\begin{abstract}
We take a new look at the details of symplectic motion in solenoid and bending magnets and rederive 
known (but not always well-known) facts. We start with a comparison of the general 
Lagrangian and Hamiltonian formalism of the harmonic oscillator and analyze the relation between
the canonical momenta and the velocities (i.e. the first derivatives of the canonical coordinates).
We show that the seemingly non-symplectic transfer maps at entrance and exit of solenoid magnets 
can be re-interpreted as transformations between the canonical and the mechanical momentum, which
differ by the vector potential.

In a second step we rederive the transfer matrix for charged particle motion in bending magnets
from the Lorentz force equation in cartesic coordinates. We rediscover the geometrical and physical 
meaning of the local {\it curvilinear} coordinate system. 
We show that analog to the case of solenoids - also the transfer matrix of bending magnets 
can be interpreted as a symplectic product of 3 non-symplectic matrices, where the entrance and
exit matrices are transformations between local cartesic and curvilinear coordinate systems.

We show that these matrices are required to compare the second moment matrices of distributions 
obtained by numerical tracking in cartesic coordinates with those that are derived by the transfer a
matrix method.
\end{abstract}

%PACS-codes:
%Cyclotrons, 29.20.dg
%Particle acceleration, classical mechanics, 45.50.Dd
%Radiation therapy equipment
%Beam optics (charged-particle beams), 41.85.-p
\pacs{41.85.-p, 45.50.Dd, 29.20.dg}
\keywords{Beam Optics, Particle Accelerators, Cyclotrons}
\maketitle

\section{Introduction}

In the course of numerical simulations of coasting beams in cyclotrons it turned out that
the eigen-emittances~\cite{eigen} computed from the second moment matrices were not constant 
as one would expect for symplectic motion~\cite{inv}. Quite obviously there was something wrong 
in the interpretation of the data. In this article we trace this error back to a missing transformation. 
The simulation tool OPAL~\cite{opal1,opal2} uses global cartesic coordinates for the integration
of the equations of motion (EQOM). The transformation to local co-moving coordinates is not
always sufficient to analyze the data properly and to compare them with second moments matrices
obtained from linear transfer matrix models like the one in Ref.~\cite{sc_paper}.
The solution of the problem might be trivial to (some) specialists, but due to the general 
context we consider it being worth a more general discussion.

The problem that we refer to, can be briefly described by either one of the following questions: 
\begin{enumerate}
\item Why are the transfer matrices at the entrance and exit of solenoid magnets considered 
to be non-symplectic~\cite{Handbook,Intro}. Is it true after all? 
\item Why is the entrance and exit of a bending magnet not considered to be non-symplectic? 
\item Is it possible to derive the transfer matrix of a bending magnet in cartesic coordinates?
\item How do we compare particle distributions generated by cartesic tracking codes with those
generated by the transfer matrix formalism?
\end{enumerate}
This work is dedicated to those readers that are not {\it ad hoc} able to give the answer to these 
questions or that are at least not sure about it.

In some sense this work is a continuation of Ref.~\cite{rdm_paper,geo_paper}, where we derived
new methods to ``solved problems'' with the general Hamiltonian of a two-dimensional harmonic 
oscillator. Here we start with the general Lagrangian description of an harmonic oscillator
and derive the Hamiltonian from it. The comparison allows us to identify the conditions
for the use of the Lagrangian state vector compared to the Hamiltonian state vector and how
they can be transformed into each other. Next we analyze the situation in case of solenoids 
and bending magnets and compare different interpretations. 
Finally we apply the resulting (simple) transformation to our numerical problem.

\section{Lagrangian of the harmonic oscillator}
\label{sec_lagrange}

In order to formulate the Lagrangian function ${\cal L}={\cal L}({\bf q},{\bf \dot q})$ of the $n$-dimensional
harmonic oscillator, we define a state vector $\phi=({\bf q},{\bf\dot q})^T=(q_1,q_2,\dots q_n,\dot q_1,\dot q_2,\dots,\dot q_n)^T$.
We then write the Lagrangian function of the harmonic oscillator in the most general way as a quadratic form:
\begeq
{\cal L}=\frac{1}{2}\,\phi^T\,{\bf L}\,\phi\,.
\endeq
The matrix ${\bf L}$ should be symmetric, as any antisymmetric component does not alter the Lagrangian function ${\cal L}$
and should therefore be physically irrelevant:
\begeq
{\bf L}=\bmtx{cc}{\bf U}&{\bf B}\\{\bf B}^T&{\bf M}\emtx\,,
\endeq
where the matrices ${\bf U}$ and ${\bf M}$ are symmetric. Written in components this is
\begeq
{\cal L}=\frac{1}{2}\,\left(q_j\,U_{jk}\,q_k+2\,q_j\,B_{jk}\,\dot q_k+\dot q_j\,M_{jk}\,\dot q_k\right)\,.
\endeq
with the $2 n\times 2 n$-matrix ${\bf L}$ and the $n\times n$-matrices ${\bf U}$, ${\bf B}$ and ${\bf M}$.
The Lagrangian equations of motion (EQOM) are:
\begeq
{d\over dt}{\partial L\over\partial\dot q_j}={\partial L\over\partial q_j}\,.
\label{eq_lagrange}
\endeq
The derivatives are explicitely:
\begary{rcl}
{\partial L\over\partial\dot q_j}&=&M_{jk}\,\dot q_k+B_{kj}\,q_k=p_j\\
{\partial L\over\partial q_j}&=&U_{jk}\,q_k+B_{jk}\,\dot q_k\\
{\partial L\over\partial{\bf\dot q}}&=&{\bf M}\,{\bf \dot q}+{\bf B}^T\,{\bf q}={\bf p}\\
{\partial L\over\partial{\bf q}}&=&{\bf U}\,{\bf q}+{\bf B}\,{\bf \dot q}\\
\endary
so that one obtains for the EQOM
\begeq
{\bf M}\,{\bf \ddot q}={\bf U}\,{\bf q}+({\bf B}-{\bf B}^T)\,{\bf \dot q}\,.
\label{eq_Leqom}
\endeq
As well-known, any matrix ${\bf B}$ can be split into two matrices ${\bf B}_s$ and ${\bf B}_a$, representing
the symmetric and the antisymmetric part:
\begary{rcl}
{\bf B}_s=({\bf B}+{\bf B}^T)/2\\
{\bf B}_a=({\bf B}-{\bf B}^T)/2\\
\endary
If one compares this with Eqn.~\ref{eq_Leqom}, one finds that the EQOM depend only on the antisymmetric (``gyroscopic'') part ${\bf B}_a$ 
while the definition of the canonical momentum includes all components of ${\bf B}$~\cite{YaSt,Talman}.
%The symmetric part of ${\bf B}$ can be interpreted as some kind of ``gauge-field''.

The number of parameters $\nu$ that can be found in the Lagrangian are the parameters that are required to 
describe two symmetric $n\times n$-matrices and an arbitrary $n\times n$-matrix:
\begeq
\nu=2\,{n\,(n+1)\over 2}+n^2=2\,n^2+n\,.
\endeq
For instance, systems with $n=2$ general degrees of freedom give $\nu=10$, for $n=3$ this gives $n=21$.
Nevertheless, with respect to the dynamics (i.e. the EQOM), some parameters can be omitted. As already
mentioned, the symmetric part of ${\bf B}$ does not enter the EQOM and secondly, the Lagrangian function
can be multiplied by an arbitrary factor without effect on the dynamics. This is a consequence of the fact
that in the Lagrangian function appears on both sides of Eqn.~\ref{eq_lagrange}, such that the any scaling
factor applied to the matrix ${\bf L}$ cancels out. However such a factor -- even though
irrelevant for the dynamics -- changes the scale of the canonical momentum:
\begeq
{\bf p}={\bf M}\,{\bf \dot q}+{\bf B}^T\,{\bf q}\,.
\label{eq_pcanonical}
\endeq

In summary one finds that the EQOM derived from the above Lagrangian contain $\nu_d$
dynamically relevant parameters. It equals the number of parameters that are required to define 
two symmetric $n\times n$-matrices and an antisymmetric $n\times n$-matrix, minus the scale factor:
\begeq
\nu_d=2\,{n\,(n+1)\over 2}+{n\,(n-1)\over 2}-1={3\,n^2+n-2\over 2}\,.
\endeq
For $n=2$ one finds $\nu_d=6$ and for $n=3$ we have $\nu_d=14$.

\section{Relation to the Hamiltonian}
\label{sec_hamilton}

The Hamilton function ${\cal H}$ is obtained by
\begary{rcl}
{\cal H}&=&p_k\,\dot q_k-{\cal L}\\
{\cal H}&=&{\bf p}^T\,{\bf \dot q}-\frac{1}{2}\,\left({\bf q}^T\,{\bf U}\,{\bf q}+{\bf q}^T\,{\bf B}\,{\bf \dot q}+{\bf \dot q}^T\,{\bf B}^T\,{\bf q}+{\bf \dot q}^T\,{\bf M}\,{\bf \dot q}\right)\\
\label{eq_Hamilton0}
\endary
We assume that the mass matrix ${\bf M}$ is invertible and replace ${\bf \dot q}={\bf M}^{-1}\,({\bf p}-{\bf B}^T\,{\bf q})$. 
If the Hamiltonian state vector $\psi$ is defined as $\psi=({\bf q},{\bf p})$, then the Hamiltonian function is derived in a few steps
\begary{rcl}
{\cal H}&=&\frac{1}{2}\,\psi^T\,{\bf H}\,\psi\\
&=&\frac{1}{2}\,\bmtx{c}{\bf q}\\{\bf p}\emtx^T\,\bmtx{cc}{\bf B}\,{\bf M}^{-1}\,{\bf B}^T-{\bf U}& -{\bf B}\,{\bf M}^{-1}\\ -{\bf M}^{-1}\,{\bf B}^T& {\bf M}^{-1}\emtx\,\bmtx{c}{\bf q}\\{\bf p}\emtx\,
\label{eq_HamiltonMatrix}
\endary
The symplectic unit matrix ${\bf\y}_0$ is given (in this representation) by
\begeq
{\bf\y}_0=\bmtx{cc}{\bf 0}&{\bf 1}\\-{\bf 1}&{\bf 0}\emtx\,,
\label{eq_sym_unit1}
\endeq
so that the Hamiltonian EQOM are
\begary{rcl}
\bmtx{c}{\bf\dot q}\\{\bf\dot p}\emtx&=&{\bf\y}_0\,\bmtx{cc}{\bf B}\,{\bf M}^{-1}\,{\bf B}^T-{\bf U}& -{\bf B}\,{\bf M}^{-1}\\ -{\bf M}^{-1}\,{\bf B}^T& {\bf M}^{-1}\emtx\,\bmtx{c}{\bf q}\\{\bf p}\emtx\\
&=&\bmtx{cc}-{\bf M}^{-1}\,{\bf B}^T& {\bf M}^{-1}\\{\bf U}-{\bf B}\,{\bf M}^{-1}\,{\bf B}^T& {\bf B}\,{\bf M}^{-1}\emtx\,\bmtx{c}{\bf q}\\{\bf p}\emtx\\
\endary
The Hamiltonian state vector $\psi$ and the Lagrangian state vector $\phi$ are related by $\psi={\bf Q}\,\phi$:
\begeq
\bmtx{c}
{\bf q}\\
{\bf p}\\
\emtx=
\bmtx{cc}
{\bf 1}&{\bf 0}\\
{\bf B}^T&{\bf M}\\
\emtx\,
\bmtx{c}
{\bf q}\\
{\bf\dot q}\\
\emtx
\label{eq_phi_psi}
\endeq
and
\begeq
\bmtx{c}
{\bf q}\\
{\bf\dot q}\\
\emtx=
\bmtx{cc}
{\bf 1}&{\bf 0}\\
-{\bf M}^{-1}\,{\bf B}^T&{\bf M}^{-1}\\
\emtx\,
\bmtx{c}
{\bf q}\\
{\bf p}\\
\emtx
\endeq
This coordinate transformation is symplectic, if
\begeq
{\bf Q}\,\y_0\,{\bf Q}^T=\y_0\,,
\endeq
or explicitely
\begary{rcl}
{\bf\y}_0&=&\bmtx{cc}
{\bf 1}&{\bf 0}\\
{\bf B}^T&{\bf M}\\
\emtx\,{\bf\y}_0\,
\bmtx{cc}
{\bf 1}&{\bf B}\\
{\bf 0}&{\bf M}\\
\emtx\\
\bmtx{cc}
{\bf 0}&{\bf 1}\\
-{\bf 1}&{\bf 0}\\
\emtx&=&\bmtx{cc}
{\bf 0}&{\bf M}\\
-{\bf M}&{\bf B}^T\,{\bf M}-{\bf M}\,{\bf B}\\
\emtx\\
\Rightarrow&&{\bf M}={\bf 1}\\
\Rightarrow&&{\bf B}^T={\bf B}\,,
\endary
i.e. it is symplectic, if (and only if) the mass matrix ${\bf M}$ equals the unit
matrix~\cite{fn1} and if ${\bf B}$ is symmetric which means that no gyroscopic forces are present.
Only in this case it is legitimate to identify ${\bf p}$ and ${\bf\dot q}$ (up to a symplectic transformation).
The first condition is usually fulfilled, if the system describes a {\it single particle with $n$ degrees
of freedom} -- instead of for example $n$ coupled particles with different masses in a linear chain.

\section{The Solenoid Magnet}
\label{sec_solenoid}

The second condition is not always fulfilled. Consider for instance the transfer-matrix ${\bf T}$ that
describes the transversal motion of a charged particle through the fringe field of a solenoid magnet~\cite{fn2}.
In the coordinate ordering used so far it is~\cite{Hinterberger,Intro}:
\begeq
{\bf T}=\bmtx{cccc}
1&0&0&0\\
0&1&0&0\\
0&\pm K&1&0\\
\mp K&0&0&1\\
\emtx\,.
\label{eq_Tsolenoid}
\endeq
This is a nice example for the transformation from $\phi$ to $\psi$ (or vice versa) with non-vanishing
gyroscopic terms. The matrices are formally {\it non-symplectic}~\cite{Handbook,Intro}, but it would be a 
misinterpretation to believe that the {\it (equation of) motion} in the fringe fields of solenoid magnets is {\it non-symplectic}. 
This is not the case. The concept of symplectic motion is based on Hamiltonian dynamics and it presumes the use of {\it canonical} momenta. 
The above transformation ${\bf T}$ is only required if one uses the state vector $\phi$ instead of $\psi$, i.e. the mechanical instead 
of the canonical momentum. If this difference is not properly taken into account, the motion {\it appears} to be non-symplectic~\cite{Intro}.

The {\it gyroscopic} terms of the matrix ${\bf B}_a$ are connected to the (derivatives of the) vector potential as 
one would expect by $\vec p_{can}=\vec p_{mech}+\vec A(\vec x)$ (using units where $q=1$ and $m=1$)~\cite{Intro}. 
In the linear 3-dimensional case one finds: 
\begary{rcl}
{\bf B}_a&=&\frac{1}{2}\,\bmtx{ccc}
0&-B_z&B_y\\
B_z&0&-B_x\\
-B_y&B_x&0\\
\emtx\\
&&\\
\vec A&=&{\bf B}\,{\bf q}={\bf B}_s\,{\bf q}+\frac{1}{2}\,\bmtx{c}
-B_z\,y+B_y\, z\\
B_z\,x-B_x\, z\\
-B_y\,x+B_x\,y\\
\emtx\,,
\endary
which directly yields
\begary{rcl}
\vec\nabla\times\vec A&=&(B_x,B_y,B_z)^T\\
&&\\
\vec\nabla\cdot\vec A&=&Tr({\bf B})=Tr({\bf B}_s)\,.
\endary
Assuming for the moment that ${\bf B}_s=0$ one finds with $K={B_z\over 2\,(B\,\rho)}$ that
the matrix ${\bf T}$ corresponds exactly to the 2-dim. transformation from $\phi$ to $\psi$ as given
in Eqn.~\ref{eq_phi_psi}. This matrix needs to be applied, since the {\it entrance} of a solenoid
is a transition from the field free region where $\psi=\phi$ to a region with gyroscopic force, where
the canonical momentum is not identical with the mechanical momentum~\cite{fn3}.

The symmetric part of ${\bf B}$ represents a symplectic transformation which is irrelevant for the 
dynamics expressed by the coordinates.
In this sense it is a similar to a ``gauge field'' that changes exclusively the canonical momentum.
The antisymmetric (``gyroscopic'') part of ${\bf B}$ is (in 3 dimensions) equivalent to the magnetic
field and one can literally identify the vector potential $\vec A$ with ${\bf B}\,{\bf q}$.

Indeed the misinterpretation of the matrices that describe the entrance and the exit of solenoids
magnets also leads to seemingly non-symplectic motion inside the solenoid magnet. The transfer matrix
$M_{sol}$ of the solenoid field is in the above coordinate ordering~\cite{Hinterberger}:
\begeq
{\bf M}_{sol}=\bmtx{cccc}
1&0&{L\over\alpha}\,S&{L\over\alpha}\,(C-1)\\
0&1&{L\over\alpha}\,(1-C)&{L\over\alpha}\,S\\
0&0&C&-S\\
0&0&S&C\\
\emtx\,,
\label{eq_Msol}
\endeq
where $S=\sin{(\alpha)}$ and $C=\cos{(\alpha)}$, which is formally also non-symplectic.
But the product of the matrix for the entrance field ${\bf T}$ (Eqn.~\ref{eq_Tsolenoid}), 
${\bf M}_{sol}$ and ${\bf T}^{-1}$ turns out to be symplectic. Hence we have:
\begeq
({\bf T}\,{\bf M}_{sol}\,{\bf T}^{-1})\,\y_0\,({\bf T}\,{\bf M}_{sol}\,{\bf T}^{-1})^T=\y_0\\
\endeq
from which one derives in a few steps:
\begary{rcl}
{\bf T}^{-1}\,\y_0\,({\bf T}^{-1})^T&=&{\bf M}_{sol}\,{\bf T}^{-1}\,\y_0\,({\bf T}^{-1})^T\,{\bf M}_{sol}^T\\
\tilde\y_0&=&{\bf M}_{sol}\,\tilde\y_0\,{\bf M}_{sol}^T\,,
\endary
so that one may also re-interpret the process as a transformation of the symplectic unit matrix:
\begeq
\tilde\y_0={\bf T}^{-1}\,\y_0\,({\bf T}^{-1})^T\,.
\endeq
But in fact, what it really describes is a change of the vector potential.

\section{Bending Magnets}
\label{sec_bending}

In the previous section we developed a proper interpretation of the matrices that describe particle motion 
at the entrance of a solenoid magnet. This raises the question, if there is an analog phenomenon
at the entrance of bending magnets. In order to clarify this, we rederive the transfer
matrix of a bending magnet in the following. Again we ignore motion parallel to the magnetic field,
which is in this case the axial (i.e. transverse vertical) motion.

Motion of charged particles in electromagnetic fields is
described by the Lorentz force equation:
\begeq
{d\vec p\over dt}=q\,(\vec E+\vec v\times\vec B)\,,
\endeq
written in cartesic coordinates:
\begary{rcl}
{d p_x\over dt}&=&q\,(E_x+v_y\,B_z-v_z\,B_y)\\
{d p_y\over dt}&=&q\,(E_y+v_z\,B_x-v_x\,B_z)\\
{d p_z\over dt}&=&q\,(E_z+v_x\,B_y-v_y\,B_x)\,.
\endary
We choose the $z$-coordinate as the vertical (axial) direction so that $x$ and $y$
and the horizontal coordinates. The motion in the median plane of a bending magnet 
is then (in the absence of acceleration) described by:
\begary{rcl}
{d p_x\over dt}&=&q\,v_y\,B_z\\
{d p_y\over dt}&=&-q\,v_x\,B_z\\
\endary
In a first step, we devide both equations by $m\,\g$, which is (in the absence of acceleration) constant:
\begary{rcl}
{d v_x\over dt}&=&{q\over m\,\g}\,v_y\,B_z\\
{d v_y\over dt}&=&-{q\over m\,\g}\,v_x\,B_z\\
\endary
We consider the {\it orbit} as the trajectory of the reference particle and 
we aim for a description of the motion in the vicinity of the orbit, i.e. of the 
{\it trajectories} of particles with small deviations from the orbit. 

We start with the state vectors of the orbit $\psi_o$ and of the trajectory $\psi$ 
in cartesic coordinates $\psi=(x,v_x,y,v_y)^T$~\cite{fn4}:
\begary{rcl}
{d\over dt}\,\psi&=&{\bf F}\,\psi=\bmtx{cccc}
0&1&0&0\\
0&0&0&{q\over m\,\g}\,B_z\\
0&0&0&1\\
0&-{q\over m\,\g}\,B_z&0&0\\
\emtx\,\psi\\
\endary
Since $B_z$ is the only relevant component in the median plane, we skip the ``z'' from now on. 
Furthermore, we like to have a mathematically positive angular velocity and hence for positive
charge we need to have a negative field $B_z$, so that we define $B=-B_z$.

A rotation in the horizontal plane is described by the following generator matrix~\cite{rdm_paper,geo_paper}:
\begary{rcl}
{\bf F}_{rot}&=&\w\,\bmtx{cccc}
0&0&-1&0\\
0&0&0&-1\\
1&0&0&0\\
0&1&0&0\\
\emtx\\
\endary
The coordinate transformation into the rotating frame is then done by subtracting the
rotational ``force matrix'' from the matrix ${\bf F}$~\cite{fn5}:
\begary{rcl}
{d\over dt}\,\psi&=&{\bf F}\,\psi=\bmtx{cccc}
0&1&\w&0\\
0&0&0&-{q\over m\,\g}\,B+\w\\
-\w&0&0&1\\
0&-\w+{q\over m\,\g}\,B&0&0\\
\emtx\,\psi\\
\endary
For synchronous rotation the rotational frequency $\w$ must equal ${q\over m\,\g}\,B$, so that 
one obtains in the co-moving frame
\begary{rcl}
{d\over dt}\,\psi&=&{\bf F}\,\psi=\bmtx{cccc}
0&1&\w&0\\
0&0&0&0\\
-\w&0&0&1\\
0&0&0&0\\
\emtx\,\psi\\
\label{eq_rot}
\endary
Next we consider small deviations from the orbit $\psi_o$ and write:
\begary{rcl}
{d\over dt}\,\psi_o&=&{\bf F}_o\,\psi_o\\
{d\over dt}\,\psi&=&{\bf F}\,\psi\\
{d\over dt}\,(\psi-\psi_o)&=&{\bf F}\,\psi-{\bf F}_o\,\psi_o\\
%&=&{\bf F}\,\psi-{\bf F}_o\,\psi+{\bf F}_o\,\psi-{\bf F}_o\,\psi_o\\
%&=&({\bf F}-{\bf F}_o)\,\psi+{\bf F}_o\,(\psi-\psi_o)\\
{d\over dt}\,\delta\psi&=&({\bf F}-{\bf F}_o)\,\psi+{\bf F}_o\,\delta\,\psi\,.
\endary
Since the condition $\w={q\over m\,\g}\,B$ holds only for the orbit (but not for all trajectories), we express
the deviations by a Taylor series which we evaluate at the orbit parameters and truncate to the linear terms:
\begary{rcl}
{1\over\g}&=&{1\over\g_o}-\g_o\,{v_o\over c^2}\,(v-v_o)={1\over\g_o}(1-\g_o^2\,{v_o^2\over c^2}\,{v-v_o\over v_o})\\
&=&{1\over\g_o}(1-\g_o^2\,{\beta_o^2}\,{\delta v\over v_o})\\
\endary
and 
\begary{rcl}
B&=&B_o+{d B\over dx}\,(x-x_o)\\
 &=&B_o\,(1+{1\over B_o}{d B\over dx}\,\delta x)\\
\endary
Note that we did not include a term with ${d B\over dy}\,\delta y$, since a field change along the longitudinal coordinate
contradicts our assumption that $\w=\mathrm{const}$.
We then find (neglecting higher order terms): 
\begeq
{q\,B\over m\,\g}\to{q\,B_o\over m\,\g}\,(1-{\g^2\,\beta^2\over v_o}\,\delta v+{1\over B_o}{d B\over dx}\,\delta x)\,.
\endeq
and hence  $(\delta\,{\bf F}={\bf F}-{\bf F}_o)$ is given by
\begeq
\delta\,{\bf F}=\bmtx{cccc}
0&0&0&0\\
0&0&0&-f\\
0&0&0&0\\
0&f&0&0\\
\emtx\,,
\endeq
where 
\begeq
f={q\,B_o\over m\,\g}\,(-{\g^2\,\beta^2\over v_o}\,\delta v+{1\over B_o}{d B\over dx}\,\delta x)\,.
\endeq
To this point we merely transformed into the rotating frame. The {\it global} coordinates of the orbit
in the rotating frame must be constant (but not necessarily zero). The time derivative of the {\it orbit}
must vanish in the rotating frame, so that we expect from Eqn.~\ref{eq_rot} 
\begeq
{\bf F}\,\psi_o=0\,,
\endeq
which is fulfilled by $\psi_o=(\rho,0,0,v_o)^T$, if
\begeq
v_o=\w\,\rho\,.
\endeq
This choice means that we choose $x$ to be the horizontal transverse and $y$ to be the longitudinal coordinate, from
which we conclude that $v_y\approx v\gg v_x$.
Then we find (again skipping higher orders)
\begeq
\delta\,{\bf F}\,\psi=\delta\,{\bf F}\,(\psi_o+\delta\psi)\approx\delta\,{\bf F}\,\psi_o\,,
\endeq
so that with $\delta\psi=(\delta x,v_x,\delta y,\delta v)^T$ one finds
\begary{rcl}
{d\over dt}\,\delta\psi&=&({\bf F}-{\bf F}_o)\,\psi_o+{\bf F}_o\,\delta\,\psi\\
&=&\bmtx{c}
0\\
-v_o\,\w\,(-{\g^2\,\beta^2\over v_o}\,\delta v+{1\over B_o}{d B\over dx}\,\delta x)\\
0\\
0\\
\emtx\\&+&\bmtx{cccc}
0&1&\w&0\\
0&0&0&0\\
-\w&0&0&1\\
0&0&0&0\\
\emtx\,\delta\,\psi\\
&=&\bmtx{cccc}
0&1&\w&0\\
-v_o\,\w\,{1\over B_o}{d B\over dx}&0&0&v_o\,\w\,{\g^2\,\beta^2\over v_o}\\
-\w&0&0&1\\
0&0&0&0\\
\emtx\,\delta\,\psi={\bf\tilde F}\,\delta\psi\\
\endary
We devide both sides by $v_o$ so that with ${d\psi\over ds}={1\over v_o}\,{d\psi\over dt}$ we obtain
\begary{rcl}
{d\over ds}\,\delta\psi&=&
\bmtx{cccc}
0&{1\over v_o}&{\w\over v_o}&0\\
-\w\,{1\over B_o}{d B\over dx}&0&0&\w\,{\g^2\,\beta^2\over v_o}\\
-{\w\over v_o}&0&0&{1\over v_o}\\
0&0&0&0\\
\emtx\,\delta\,\psi\\
\endary
In the following we apply a sequence of 3 transformations described by matrices ${\bf T}_i$, 
where each transformation is of the general form
\begary{rcl}
\delta\psi&\to&{\bf T}_i\,\delta\,\psi\\
{\bf F}&\to&{\bf T}_i\,{\bf F}\,{\bf T}_i^{-1}\,,
\endary
where we omitted the tilde of the force matrix ${\bf F}$ for a better readability.

The first transformation matrix ${\bf T}_1$ is used to scale the velocities by $1/v_o$ and is 
given by:
\begeq
{\bf T}_1=\mathrm{Diag}(1,{1\over v_o},1,{1\over v_o})\,,
\endeq
so that 
\begeq
{\bf F}=
\bmtx{cccc}
0&1&{\w\over v_o}&0\\
-{\w\over v_o}\,{1\over B_o}{d B\over dx}&0&0&{\w\over v_o}\,\g^2\,\beta^2\\
-{\w\over v_o}&0&0&1\\
0&0&0&0\\
\emtx\,,
\endeq
and hence $\delta\psi$ is now given by:
\begeq
\delta\psi=(\delta x,{\delta v_x\over v_o},\delta y,{\delta v\over v_o})^T\,.
\endeq
Due to the choice of $\psi_o=(x_o,0,0,v_o)^T$, $\delta x$ is the local horizontal, $\delta y=y$ the local longitudinal coordinate and
${\delta v_y\over v_o}\approx{\delta v\over v_o }={1\over\g^2}\,{\delta p\over p}$ is the velocity deviation, so that with the field index
$n_x$ define by $n_x={\rho\over B_o}{d B\over dx}$ and ${w\over v_o}={1\over\rho}$ we obtain
\begeq
{\bf F}=
\bmtx{cccc}
0&1&{1\over\rho}&0\\
-{n_x\over\rho^2}&0&0&{\g^2\,\beta^2\over\rho}\\
-{1\over\rho}&0&0&1\\
0&0&0&0\\
\emtx\,.
\endeq
Next we transform from the velocity deviation to the momentum deviation using ${\bf T}_2$
\begeq
{\bf T}_2=\mathrm{Diag}(1,1,1,\y^2)\,.
\endeq
The result is:
\begeq
{\bf F}=
\bmtx{cccc}
0&1&{1\over\rho}&0\\
-{n_x\over\rho}&0&0&{\beta^2\over\rho}\\
-{1\over\rho}&0&0&{1\over\y^2}\\
0&0&0&0\\
\emtx\,.
\label{eq_Fb}
\endeq
\begin{figure}
\includegraphics[width=8cm]{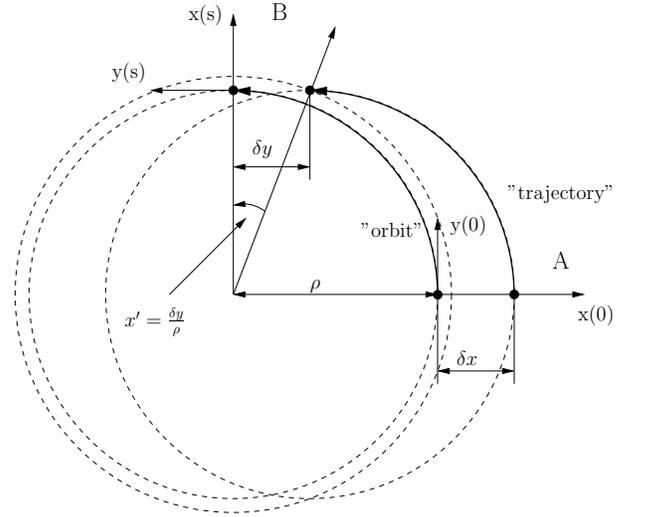}
\caption[]{
Transformation into curvilinear coordinate system. The trajectory with deviation $\delta x$ in position A
causes a deviation $\delta y$ in position B, where one finds no direction difference in cartesic coordinates.
Interpreted in curvilinear (i.e. cylindrical) coordinates one has (in first order) a direction deviation 
$x'={\delta y\over\rho}$.
\label{fig_curvilin}}
\end{figure}
The last transformation  ${\bf T}_3$ required to obtain the well-known transfer matrix of a bending magnet, transforms from 
the local co-moving cartesic system to the local co-moving {\it curvilinear} system. The transformation is explained in 
Fig.~\ref{fig_curvilin}:
\begeq
{\bf T}_3=\bmtx{cccc}
1&0&0&0\\
0&1&{1\over\rho}&0\\
0&0&1&0\\
0&0&0&1\\
\emtx\,.
\label{eq_T3}
\endeq
This last transformation yields finally:
\begeq
{\bf F}=\bmtx{cccc}
0&1&0&0\\
-{1+n_x\over\rho^2}&0&0&{1\over\rho}\\
-{1\over\rho}&0&0&{1\over\g^2}\\
0&0&0&0\\
\emtx\,.
\endeq
The (symplectic) transfer matrix ${\bf M}_b=\exp{({\bf F}\,s)}$ then is:
{\small\begary{rcl}
{\bf M}_b&=&\bmtx{cccc}
C&{\rho\,S\over\sqrt{1+n_x}}&0&\rho\,{1-C\over 1+n_x}\\
-{\sqrt{1+n_x}\over\rho}\,S&C&0&{S\over\sqrt{1+n_x}}\\
-{S\over\sqrt{1+n_x}}&-{\rho\,(1-C)\over 1+n_x}&1&{\rho\,S\over(1+n_x)^{3/2}}+{s\over\g^2}-{s\over 1+n_x}\\
0&0&0&1\\
\emtx\\
S&=&\sin{(\alpha\,\sqrt{1+n_x})}\\
C&=&\cos{(\alpha\,\sqrt{1+n_x})}\,,
\label{eq_Mb}
\endary}
where the bending angle $\alpha$ is given by $\alpha={s\over\rho}$. 
As in case of the solenoid magnet, it is possible to split the transfer matrix ${\bf M}_b$ into 3 parts, first the 
transformation into curvilinear coordinates ${\bf T}_3$ which then represents the fringe field 
(without entrance angle), second the transfer matrix of the bending magnet ``itsself''
and finally the transformation ${\bf T}_3^{-1}$ back to cartesic coordinates. 
The transfer matrix for the bending magnet (analog to ${\bf M}_{sol}$ as given in Eqn.~\ref{eq_Msol}) 
is the matrix exponent of the force matrix (as given by Eqn.~\ref{eq_Fb}) multiplied by the pathlength 
$s=\alpha\,\rho$ and is explicitely given by:
{\small\begary{rcl}
{\bf M}_{bend}&=&\exp{({\bf F}\,s)}\\
&=&\bmtx{cccc}
C&{\rho\,S\over\sqrt{k}}&{S\over\sqrt{k}}&{\rho\,(1-C)\over k}\\
-{n_x\,S\over\rho\,\sqrt{k}}&{1+n_x\,C\over k}&{(C-1)\,n_x\over \rho\,k}&X\\
-{S\over\sqrt{k}}&{\rho\,(C-1)\over k}&{C+n_x\over k}&Y\\
0&0&0&1\\
\emtx\\
X&=&{\alpha\,(\y^2-1)\,\sqrt{k}+n_x\,(\y^2\,S-\alpha\,\sqrt{k})\over\y^2\,k^{3/2}}\\
Y&=&\rho\,\left({S\over k^{3/2}}+\alpha\,({1\over\y^2}-{1\over k})\right)\\
k&=&1+n_x\\
S&=&\sin{(\alpha\,\sqrt{k})}\\
C&=&\cos{(\alpha\,\sqrt{k})}\,,
\endary}
where $\alpha$ is the bending angle of the magnet and $\rho$ the 
bending radius or the orbit.
Then one verifies from Eqn.~\ref{eq_T3} and Eqn.~\ref{eq_Mb}:
\begeq
{\bf M}_b={\bf T}_3\,{\bf M}_{bend}\,{\bf T}_3^{-1}\,,
\endeq
so that the complete symplectic transfer matrix of a bending magnet may be regarded
as a product of 3 ``non-symplectic'' matrices, just as one finds it for solenoids. 
In essence we merely applied the equation 
\begeq
\exp{({\bf T}_3\,{\bf F}\,{\bf T}_3^{-1}\,s)}={\bf T}_3\,\exp{({\bf F}\,s)}\,{\bf T}_3^{-1}\,,
\endeq
which we believe to reflect the essential difference in typical textbook descriptions of 
bending magnets (left side, symplectic) and solenoids (right side, 3 times ``non-symplectic'').

In order to facilitate comparison with Sec.~\ref{sec_solenoid}, we go back to the
coordinate ordering from Sec.~\ref{sec_lagrange}, i.e. first the coordinates
and then the momenta (or ``velocities''). The matrix ${\bf T}_3$ is then written as
\begeq
{\bf T}_3=\bmtx{cccc}
1&0&0&0\\
0&1&0&0\\
0&{1\over\rho}&1&0\\
0&0&0&1\\
\emtx
\endeq
If we compare this with Eqn.~\ref{eq_Tsolenoid} and Eqn.~\ref{eq_phi_psi}, then
we find that the difference is merely the {\it gauge} represented by a symmetric
matrix ${\bf B}$ of the form
\begeq
{\bf B}_s=\frac{1}{2}\,\bmtx{cc}
0&-{1\over\rho}\\
-{1\over\rho}&0\\
\emtx\,.
\endeq
And as we derived above, a non-vanishing symmetric part of ${\bf B}$ equals 
a symplectic gauge-transformation without influence on the dynamics of
${\bf q}$ and ${\bf\dot q}$.

\section{Application to Numerical Tracking Computations}
\label{sec_application}

All the above developed formalism stays academic as long as we do not
refer to a practical ``problem''. In Ref.~\cite{sc_paper} we described an
iterative method to determine the parameters of a matched beam matrix of second
moments $\sigma$ for cyclotrons with strong space charge forces. Using samples
with typically $10^5$ particles~\cite{random}, the parallel framework OPAL has been used to 
simulate coasting beams in cyclotrons similar to the PSI ring machine~\cite{Ring}
and some results have been presented~\cite{scopal}. 
\begin{figure}
\includegraphics[width=8cm]{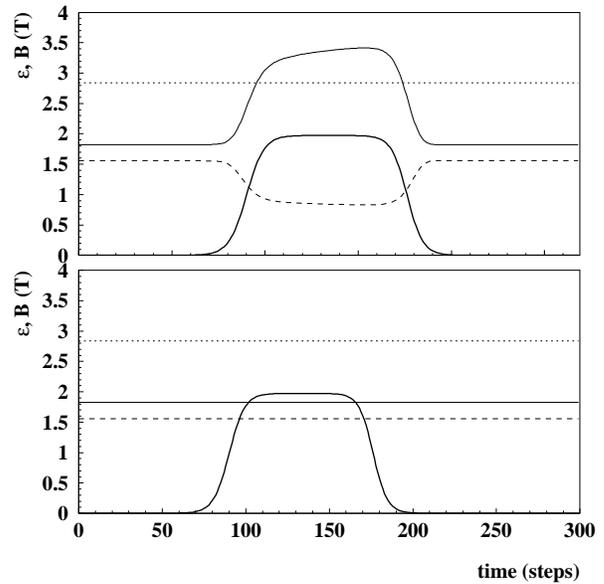}
\caption[]{
Upper graph:
Eigenvalues of the matrix ${\bf S}=\sigma\,\y_0$ of a Gaussian particle distribution tracked along the equilibrium orbit
over one sector of a separate sector ring cyclotron. The transformation ${\bf T}_3$ has not been applied.
The horizontal eigenvalues (thin solid line), the longitudinal eigenvalue (dashed line) and the product
of both (dotted line). The thick solid line shows the magnetic field in Tesla.
Lower graph: The same figure after the transformation ${\bf T}_3$. The eigenvalues are all constant
along the orbit as expected for symplectic motion.
\label{fig_opal}}
\end{figure}
The distributions turned out to be properly matched only for a starting position 
in the field free region (i.e. between sector magnets), while the matching failed 
when the tracking started somewhere within the sector magnet. 
A detailled analysis (including a cross check with a second tracking code without 
space charge solver) suggested, that the eigen-emittances from the distributions 
evaluated in cartesic coordinates where constant only in constant 
field regions, but changed from valley to sector (and vice versa).
The transformation from the local cartesic to the local curvilinear coordinate system
with the matrix ${\bf T}_3$ as derived above solved the problem and verified that the 
motion is indeed symplectic. The eigen-emittances evaluated in local cartesic and 
local curvilinear coordinate systems are shown in 
Fig.~\ref{fig_opal} as a function of time (i.e. step-number).

\section{Summary}

We investigated symplectic motion in magnetic fields using the examples of solenoid and 
bending magnets. We rederived the transfer matrix of a bending magnet starting from
the Lorentz force equation in cartesic coordinates. We found that the motion is symplectic 
in both types of magnets, if one takes the proper canonical momentum into account. 
Furthermore it turned out that there is no essential difference between solenoid and bending magnets, 
despite the fact that they are often described differently. We also found that the {\it curvature}
($1/\rho$) of the local coordinate system is intimately connected to the vector potential which is (in linear
approximation) given by the matrix ${\bf B}$ multiplied by the coordinates ${\bf q}$.

We applied these findings to tracking of particle distributions in cartesic coordinates and
gave the transformation between local curvilinear and local cartesic coordinates. 
We showed that the motion is formally symplectic only in local curvilinear coordinates.

\section{Acknowledgements}

We thank J.J. Yang for fruitful discussions about particle tracking with OPAL.
\newline
%\begin{appendix}
%\end{appendix}

\section*{References}

\bibliographystyle{model1a-num-names}

\begin{thebibliography}{00}
\bibitem{eigen} The eigen-emittances are the eigenvalues of ${\bf S}=\sigma\,\y_0$, where $\sigma$ is
the matrix of second moments and $\y_0$ is the symplectic unit matrix.
\bibitem{fn1} If one takes into account that the Lagrangian allows for a scaling factor,
the mass matrix effectivly has to be proportional to a unit matrix.
\bibitem{fn2} We ignore the motion parallel to the magnetic field, which is in case of a solenoid the longitudinal coordinate.
\bibitem{fn4} Here we choose a different coordinate ordering compared to Sec.~\ref{sec_lagrange} in order to 
facilitate comparison with the conventional notation and we postpone the question, if the state vector is Hamiltonian
of Lagrangian.
\bibitem{fn5} In analogy to Einstein's equivalence principle of a uniformly accelerated reference frame and the force of gravitation, 
a frame rotating at constant angular velocity is equivalent to a gyroscopic force, i.e. a ``magnetic'' field.
\end{thebibliography}

\end{document}